\documentclass[referee]{raa}            % referee version: for submission

\usepackage{graphicx,times}             %for PS/EPS graphics inclusion, new
\usepackage{natbib}\usepackage{color}

\newcommand{\etal}{{\it et al.}}

\begin{document}

\title{Local sunspot oscillations and umbral dots
}

\volnopage{Vol.0 (200x) No.0, 000--000}      %%preserved for Editor. DOn't remove!
\setcounter{page}{1}          %%starting page, preserved for Editor. DOn't remove!

\author{Y. Zhugzhda
\inst{1}
\and R. Sych
\inst{2}
   }

\institute{IZMIRAN,Moscow 142092, Russia; {\it YZhugzhda@mail.ru}\\
\and
Institute of Solar-Terrestrial Physics SB RAN, Irkutsk, Russia\\
   }

\date{Received~~2009 month day; accepted~~2009~~month day}

\abstract{ Data analysis of sunspot oscillation  based on 6-hr SDO run of observation showed that low frequency  ($0.2<\omega<1$mHz) oscillations are local similar to three and five minute oscillations. The oscillations in the sunspot are concentrated in cells of a few arcsec, each of which has its own oscillation spectrum. The analysis of two  scenario for sunspot oscillations leads to conclusion that local sunspot oscillations occur due to subphotospheric resonator for slow mhd waves. Empirical models of sunspot atmosphere and the theory of slow waves in thin magnetic flux tubes is applied to the modeling of subphotospheric resonator.  Spectrum of local oscillations consists of a great number of lines. This kind of spectrum can occur only if the subphospheric resonator is a magnetic tube with a rather weak magnetic field. Magnetic tubes of this sort are umbral dots that appear due to the convective tongues in the monolithic sunspots. The interrelation of local oscillations with umbral dots and wave fronts of traveling waves in sunspots is discussed.
}

\authorrunning{Y. Zhugzhda \& R. Sych}            %author_head in even pages
\titlerunning{Local oscillations and umbral dots }  % title_head in odd pages

   \maketitle
%% The author head (on even pages) and the title head (on odd pages) will be
%% automatically extracted from \author{} and \title{}. Whenever the title is too long,
%% you will be asked to supply a shorter one by inserting either \authorrunning{} or
%% \titlerunning{} before \maketitle. Anyway, you can specify your own heads.
%%
%%
%% Note: In the following text body of your manuscript, please note several differences from
%%       other major journals:
%% (1) \subsection{Please Capitalize the First Letter of Each Notional Word in Subsection Title}
%% (2) Please Capitalize the First Letter of Each Notional Word in all tables' captions

%
%________________________________________________ sections below
%
\section{Introduction}           %% first-level sections will be auto-capitalized
\label{sect:intro}
Sunspot oscillations were discovered by \citet{Beckers72}. It was so-called three minute oscillations. Sunspot oscillations are observed in photosphere, chromosphere and corona. Periods of sunspot oscillations are from minutes to hours. An enormous observational data have been stored. Ground-based and space observations are used to explore sunspot oscillations.  Until recently, those were mainly ground-based observations (see reviews \citet{Bogdan06}, and \citet{Khomenko15}). Ground-based observations allow us to use many different methods of investigation of oscillations but typically suffer from a lack of frequency resolution due to the short time series of observations. Following the SDO launch, hours-long observations of sunspot oscillations became available. These observations, along with the Nobeyama radioheliograph data, augmented the stock of knowledge of the oscillation properties (\citealt{Sych16}).

The sunspot oscillation spectrum is very wide. It comprises a great number of lines. The spectrum of the so-called three-minute oscillations involves tens of lines (\citealt{Reznikova12,Zhugzhda14}). Besides, there are five-minute sunspot oscillations, long-period ($0.2<\omega<1$mHz) and super long-period ($\omega<0.2$mHz) oscillations (\citealt{Bakunina13}). Another relevant feature of sunspot oscillations is their spatial non-uniformity.
\cite {Centeno05} revealed that the waves propagate only inside channels of subarcsecond width whereas the rest of the umbra remains nearly at rest. Numerous observations show that oscillations are localized in small  spots of umbra (\citealt {Jess12, Zhugzhda14, Prasad15, Chae17, Sych16}). We call them {\it local oscillations}. Besides, there are {\it non-local oscillations} in sunspots. Namely,  there are traveling waves that propagate through the sunspot umbra to penumbra (\citealt{Zirin72, Giovanelli72, Sych14,Su16}).  \citet{Jess17} claim that there are a non-local three-minute oscillations  with $m = 1$ slow magneto-acoustic mode.

All this places a challenge before the theory that should give explanation both local and nonlocal oscillations of sunspots. Indeed, there are two approaches to solve this problem. One approach is related to the assumption about the occurrence of resonance phenomena in the sunspot atmosphere. Within this approach, the oscillation spectrum is considered to be the spectrum of some resonators in the sunspot atmosphere. Within the resonance approach, the chromospheric resonator had been addressed for years (\citealt{Zhugzhda81, Zhugzhda08, Botha11}), because it predicted a rather complex spectrum of three-minute oscillations. However, the real spectrum appeared to be much more complex (\citealt{Reznikova12, Zhugzhda14}), than it could be explained within the chromospheric resonator model. The other approach is based on the fact that a sunspot is surrounded by a quiet solar atmosphere, in which there are p-mode oscillations of different frequencies. These oscillations should penetrate into a sunspot and convert into magnetogravity waves. In this case, the sunspot oscillation spectrum is determined by the spectrum of the arriving oscillations. To study the effect of p-mode oscillations on sunspot oscillations great efforts have been undertaken in this research area (\citealt{Khomenko15, Felipe17}). Additionally, there are super-low frequency sunspot oscillations (\citealt{Chorley10, Bakunina13}) having absolutely another nature, than that discussed above (\citealt{Solovev14}).

\citet{Zhugzhda14} studied three-minute sunspot oscillations that are known to represent slow MHD waves. Addressed were two scenarios for the origin of slow oscillations as a result of wave arrival from sunspot subphotospheric layers.  Non-local sunspot oscillations appeared to be excited by p-modes, whereas  local oscillations may be excited only by slow waves arriving from  sunspot subphotospheric layers. Thereby, the assumption that local and non-local oscillations have different natures enable to unite the above two approaches to the oscillation theory.

\citet{Zhugzhda14} arrived at the conclusion that a slow-wave subphotosphere resonator may be the source of the slow waves responsible for local three-minute oscillations. From this hypothesis, follow some conclusions that are supposed to be analyzed in this paper. The spectrum of local three-minute oscillations represents a set of a larger number of closely spaced lines. Such a spectrum may originate, as long as the subphotospheric resonator fundamental frequency is sufficiently low. In this case, the spectrum of three-minute oscillations should comprise upper oscillation harmonics. Besides, one should observe not only three-minute local oscillations, but also low-frequency local oscillations. This paper intends to prove the an existence of such oscillations.

Another important aspect is to build a simplest numerical model for the slow-wave subphotospheric resonator. The presence of the localized sunspot oscillations should be related to some fine-scale structures in the sunspot atmosphere. An obvious candidate for this role is umbral dots. A separate section of this paper compares the parameters of our simplest model for the resonator with the parameters of convective  tongues that, according to the present-day ideas, are responsible for the origin of umbral dots.
\begin{figure}    %%%%%%%%%%%%%%%%%% FIGURE 1
   \centerline{\includegraphics[width=0.8\textwidth,clip=]{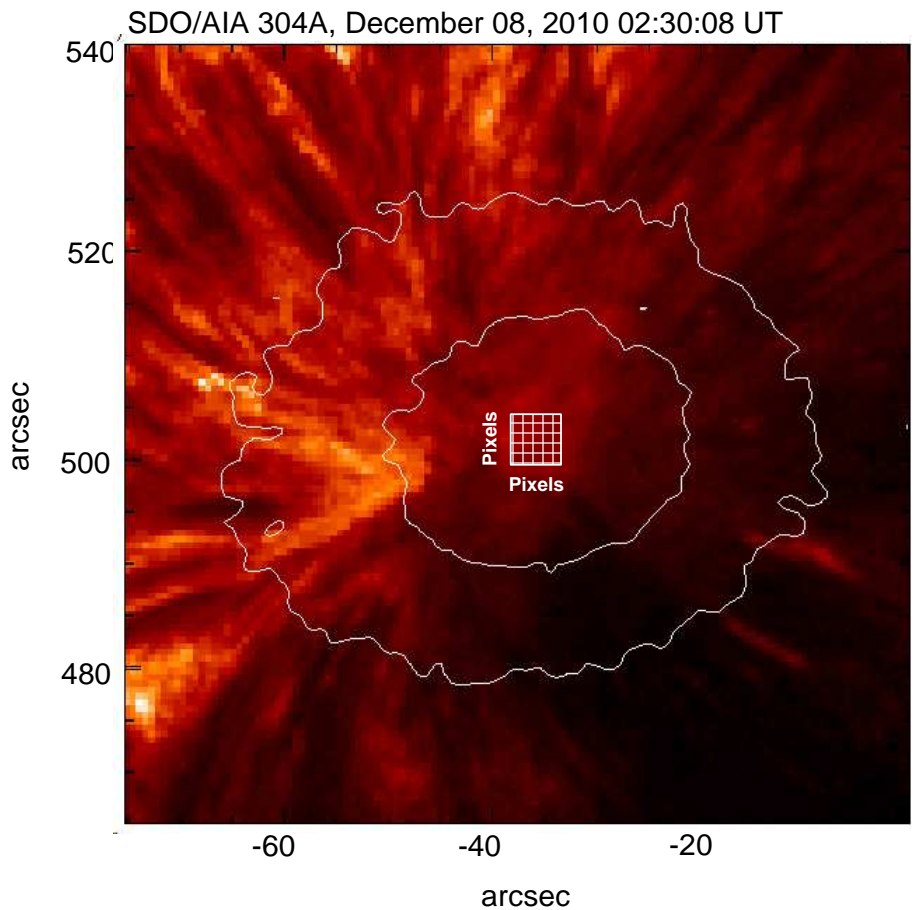}}
              \caption{The SDO/AIA $304$\AA\hspace{1pt}  sunspot image in ultraviolet range, at 2010 December 8, 02:30 UT. The contours mark the umbra and penumbra boundaries in white light. The squared site ($9\times 9$ pixel) under study is shown in the sunspot center. The pixel (1,1) is in the left bottom corner of the squared site. Coordinates are in arcsec.}
\label{fig1}
\end{figure}
Besides, we considered necessary to briefly address the phenomenon of sunspot wavefronts that, by the proposed classification, are a typical representative of non-local oscillations that should be excited by p-modes. Indeed, this is corroborated by simulations (\citealt {Felipe17}). However, these simulations and observations (\citealt{Chmielewski16,  Zhugzhda14, Sych10}) show that the sources of wave fronts should be local. Thereby, there begs a question, whether local oscillations may be the source of divergent sunspot wavefronts. We discuss this possibility in the final section of the paper.

The paper is organized as follows. First of all we show that the low-frequency and the five-minute oscillations  like a three-minute are local oscillations. Then scenarios of there-minute oscillations are generalized to the case of low-frequency oscillations. The theory of slow waves in thin flux tubes is described. Empirical models of sunspot atmosphere is used to develop a simple model of subphotospheric resonator for slow waves. The connection of local oscillations with umbral dots and wave fronts is discussed.

\section{Observations}
\label{sect:Obs}
For our analysis of local oscillations in sunspots, we selected a single, symmetric sunspot from active  region NOAA 11131. This group was passing on the solar disk in its northern hemisphere through December 2010. We studied the 2010 December 8 00:00-06:00 UT time period, when the group was crossing the central meridian. A time cube of the sunspot images obtained with the SDO/AIA in UV at the upper chromosphere / transition region level (HeII line, 304\AA, T=80000K) was used. The pixel size was 0.6 arcsec. The data were obtained at the 12-sec cadence. The observation duration was 6 hours, which allowed us to study oscillations within the 0.5 - 120 minute period range. Figure \ref{fig1} shows the sunspot at 02:30UT, with the umbra and penumbra boundaries in the white light superimposed like contours. The umbra size was 25 arcsec, the penumbra being 50 arcsec. The brightness is presented in the logarithmic scale. To obtain the images, we used the SDO/AIA http//www.lmsal.com/get\_aia\_data/ resource that allows us to obtain Lev1 calibrated images for different wavelengths within the given time period. The source selection was manual by assigning the active region center coordinates and the site: width and height in arcsec. The differential rotation of the given object during the observational time was removed by introducing an integer shift through the algorithm implemented at the website.  The data consist of 1800 images with time lapse 12 seconds.  The square site 9x9 pixel located in the sunspot center was explored. The temporal variations in the UV brightness were studied separately for each pixel of this square site.
\section{Properties of sunspot oscillations}
Our analysis of the observational data is aimed to study whether sunspot oscillations are local or not.  These oscillation properties, in our view, are key to the sunspot oscillation theory.

\subsection{Properties of oscillation spectral lines}
\begin{figure}%%%%%%%%%%%%%%%%%% FIGURE 2
\centerline{\includegraphics[width=1.0\textwidth,clip=]{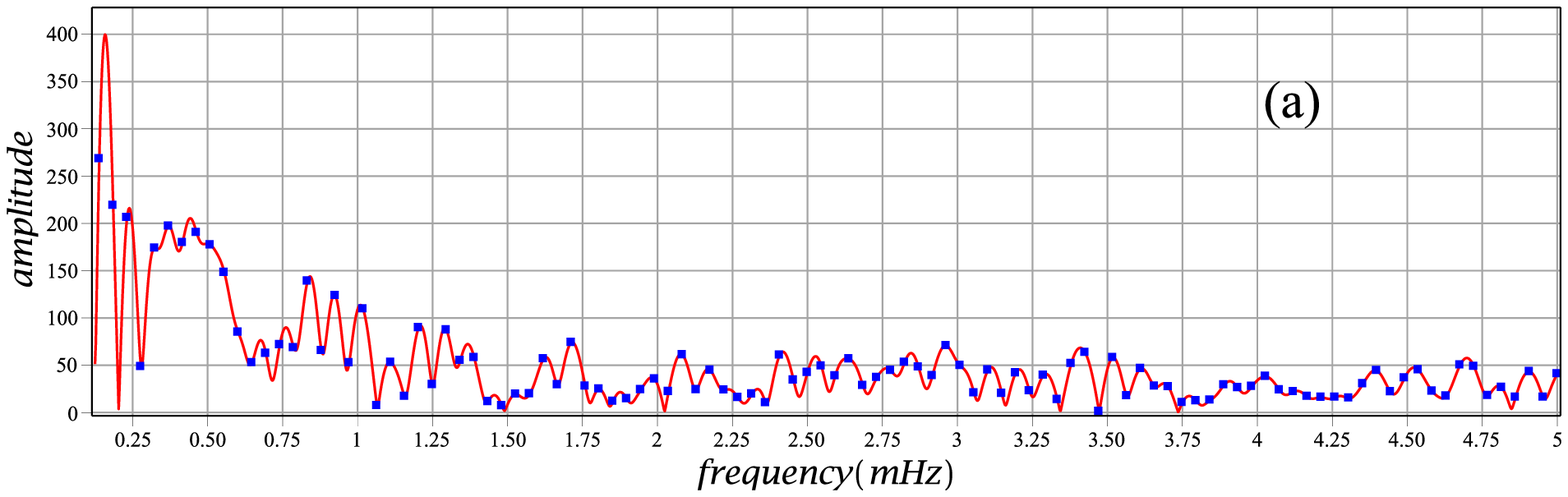}}
\centerline{\includegraphics[width=1.0\textwidth,clip=]{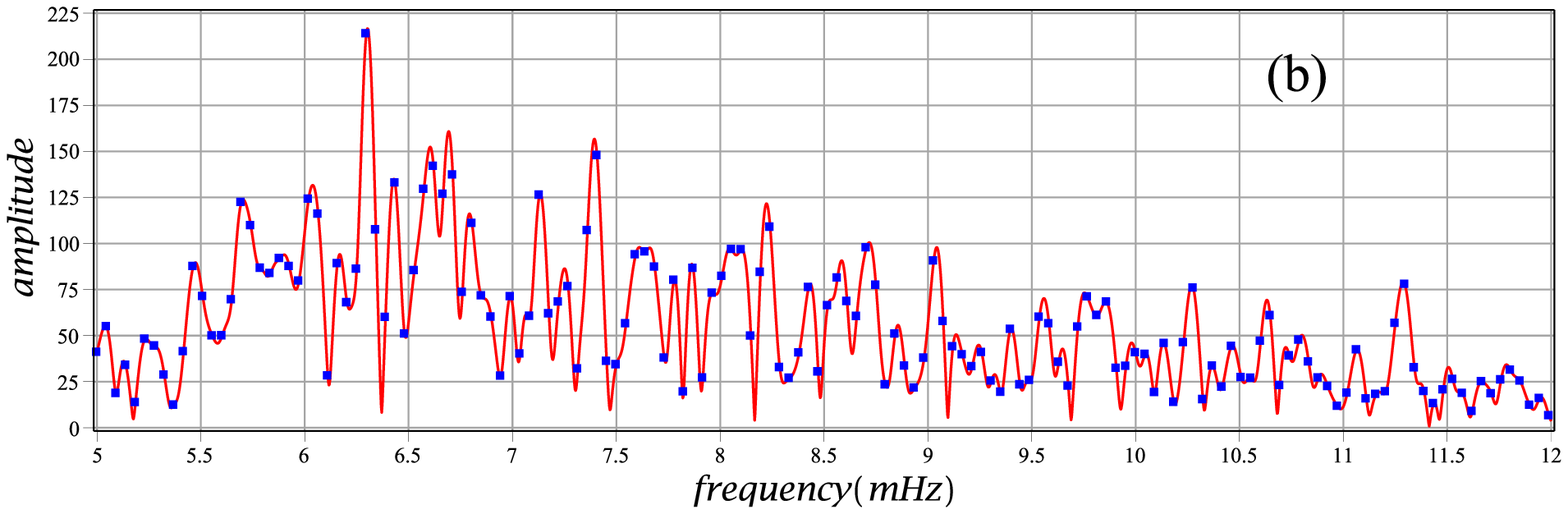}}
\centerline{\includegraphics[width=.50\textwidth,clip=]{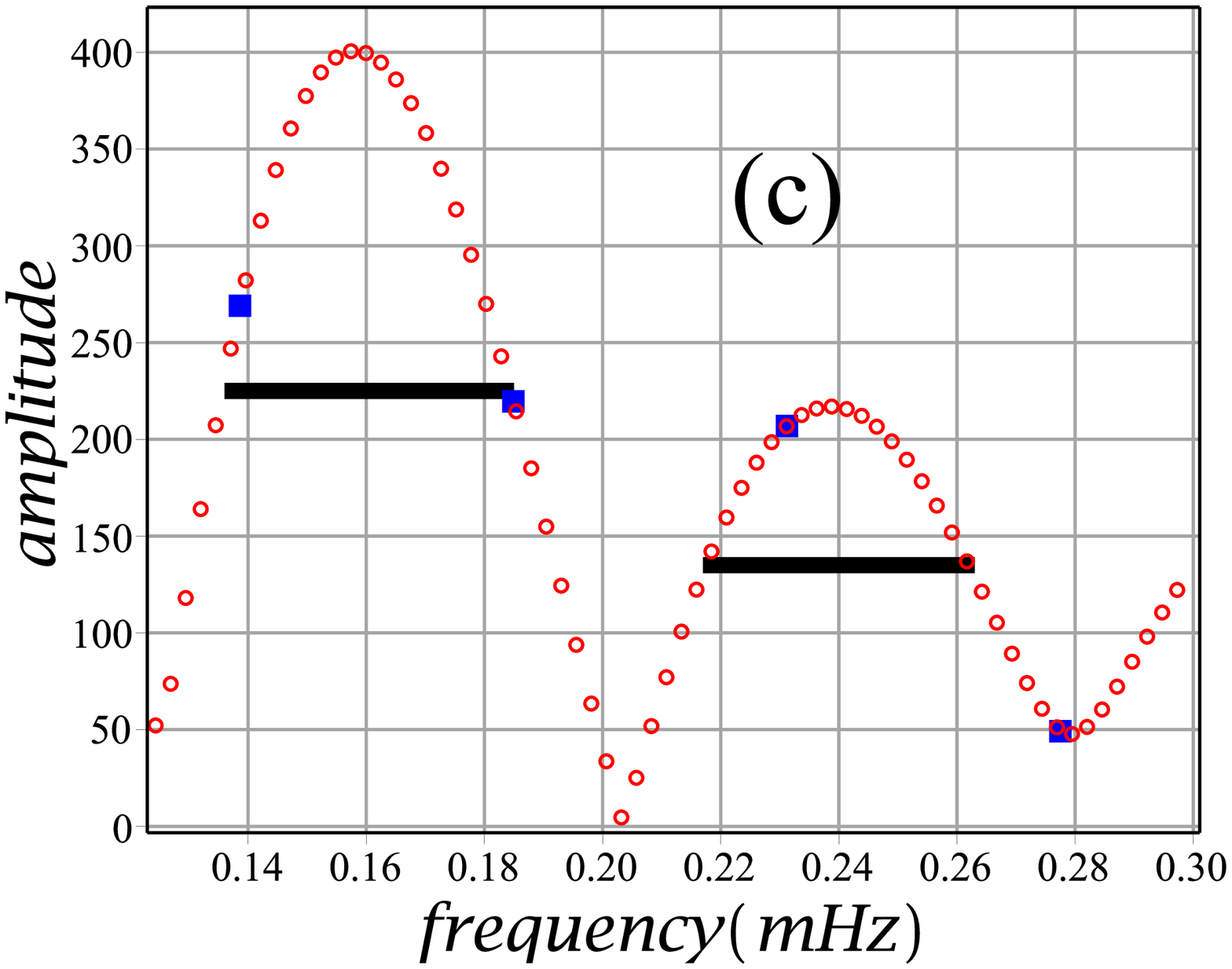}
\includegraphics[width=.50\textwidth,clip=]{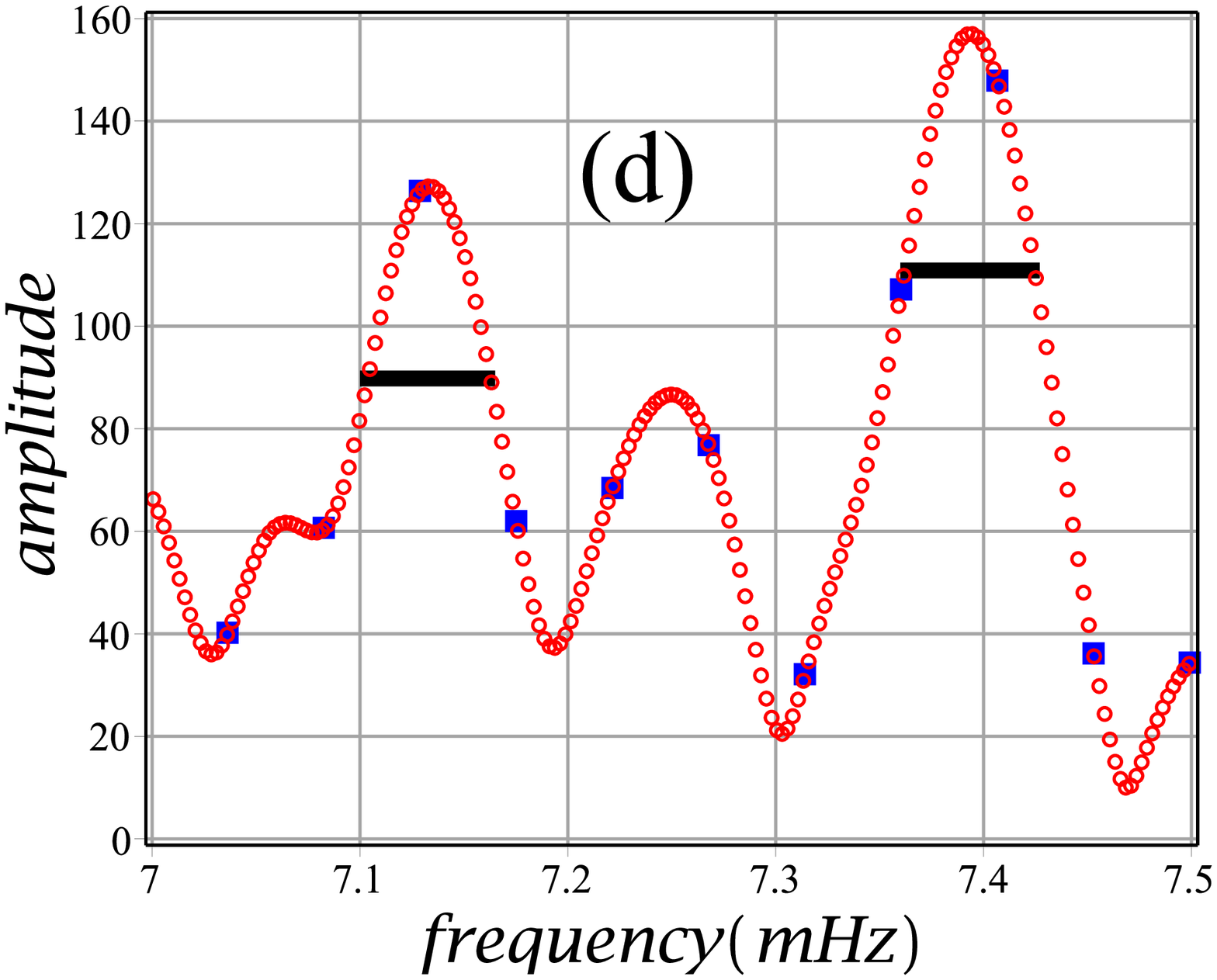}  }
\caption{ The amplitude spectrum of the pixel under study site with the coordinates (3,4) calculated without (blue fill boxes) and with (red solid line on (a,b) and circles on (c,d)) zero padding. Horizontal lines on (c,d) show the half-width of the spectral lines. }
\label{fig2}
\end{figure}

Plots of Figure \ref{fig2}a,b plot present the amplitude spectrum for  oscillations in the pixel under study site with coordinates (3,4) where, within the investigated region, the most-amplitude oscillations are observed in the sunspot center. In order to obtain spectrum of oscillations the Fourier transformation with and without the  zero padding were used. Blue squares on the graphs of Figure\ref{fig2} are the amplitude spectrum obtained by applying fast Fourier transform to the original time series after zero padding up to $2^{11}$. Spectral lines in the spectrum appear so narrow that each line accounts for only two or three squares on plots of Figure \ref{fig2} which makes it impossible to reproduce the line profile properly and to determine the position of the line maximum. To overcome this drawback the time series was extended from 1800 points to $2^{15}$ by zero padding and the fast Fourier is applied to the extended time series. Results are shown by red lines on Figure \ref{fig2}a,b and by red circles on Figure \ref{fig2}c,d.

The spectrum of  oscillations comprises  of dozens closely located spectral lines. This property of the spectrum became obvious due to a long time series, which provided a high spectral resolution. Former ground-based observations about one-hour long were enable to resolve only few lines in the oscillation spectrum of sunspot oscillations.
Figure \ref{fig2}c,d shows two small segments of the amplitude spectrum, including few spectral lines each. The Figure \ref{fig2}c,d plots show that the spectral lines are very narrow which is essential to the theory of oscillations. These plots allow us to determine the half-width of the spectral line at a high accuracy. Because the amplitude spectra are presented in these plots, the half-width should be determined by the line width at Level $1/\sqrt{2}$. The half-width of the lines appears about $\sim 0.03-0.05$ mHz, which corresponds to the characteristic time of the oscillation damping $\sim 9$ hours that is about the observational time (6 hours). In the case of three-minute oscillations the half-width corresponds to very high Q-factor, $Q\geq 500$. This seriously constraints on the theory of sunspot oscillation model. Only resonator may be responsible for the occurrence of the spectrum with such narrow lines. Lines in the low-frequency part of the spectrum do not correspond to so high Q-factor. But we expect that this is the result of insufficient spectral resolution.
\begin{figure}%%%%%%%%%%%%%%%%%% FIGURE 3
\centerline{\includegraphics[width=1.0\textwidth,clip=]{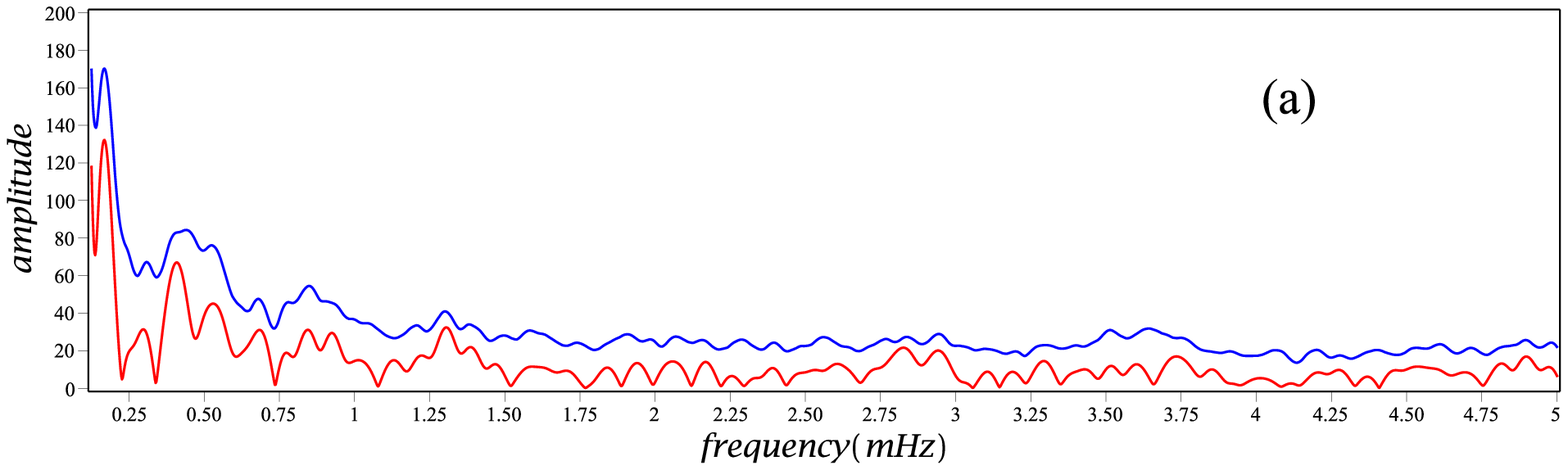}}
\centerline{\includegraphics[width=1.0\textwidth,clip=]{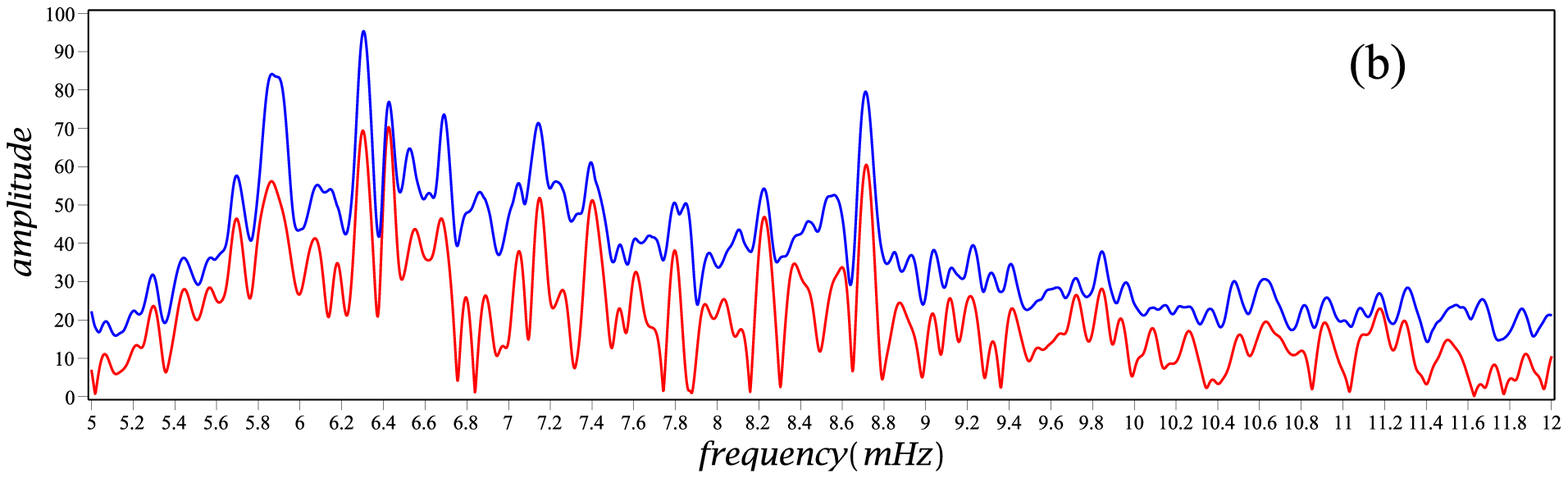}}
\centerline{\includegraphics[width=1.0\textwidth,clip=]{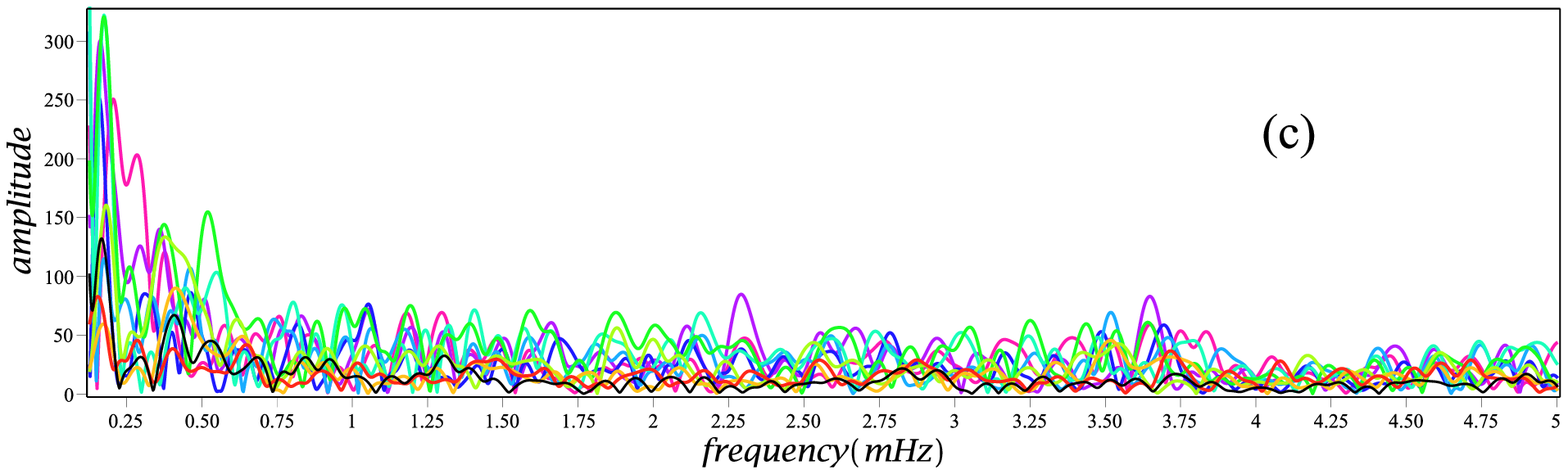}}
\centerline{\includegraphics[width=1.0\textwidth,clip=]{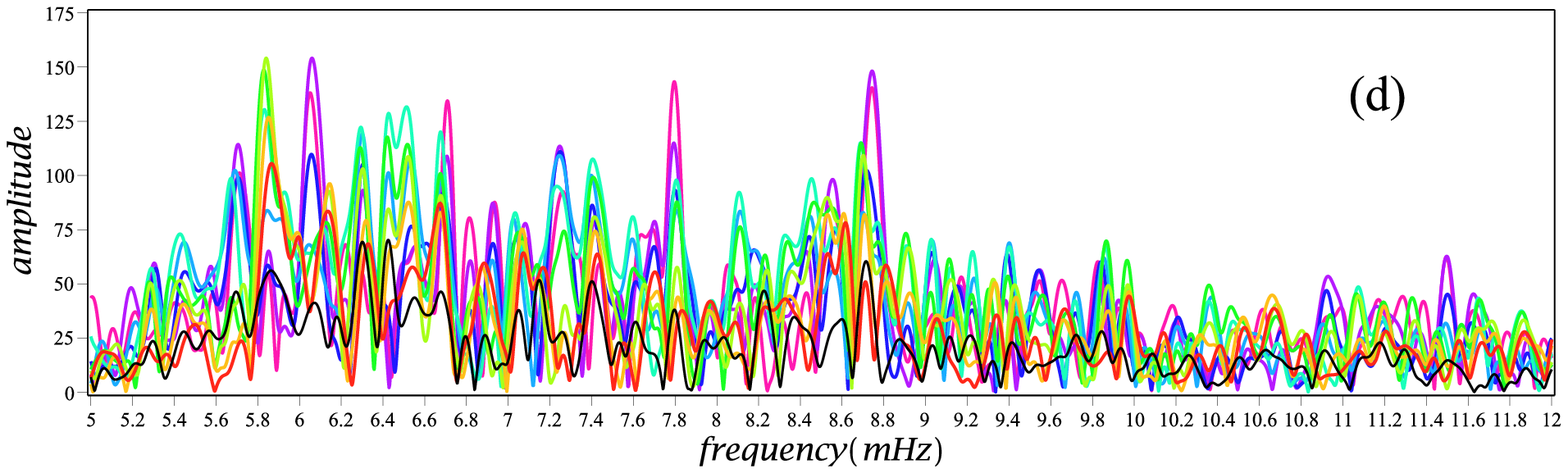}}
\caption{Amplitude spectra of sunspot oscillations. (a,b) Averaged amplitude spectrum of the site under study (blue curve) and the  amplitude spectrum of averaged oscillations of the site (red curve). (c,d) Amplitude spectra of pixels with coordinates $x=1\ldots 9,y=4$ and $x=1\ldots 9,y=6$ are presented in different colors, respectively. The black line is the same as the red line on the graphs (a, b). }\label{fig3}
\end{figure}
\subsection{Locality of sunspot oscillations}
Another key issue for the oscillation theory is whether sunspot oscillations are local or non-local, i.e., a sunspot oscillates as a whole, or those are broken down into individual cells which oscillate more or less independently of each other. \citet{Zhugzhda14} have revealed that 3 minute oscillations are local oscillations. The goal of current exploration is to check whether the locality is a general property of sunspot oscillations.
\begin{figure}%%%%%%%%%%%%%%%%%% FIGURE 4
 \centerline{\includegraphics[width=1.0\textwidth,clip=]{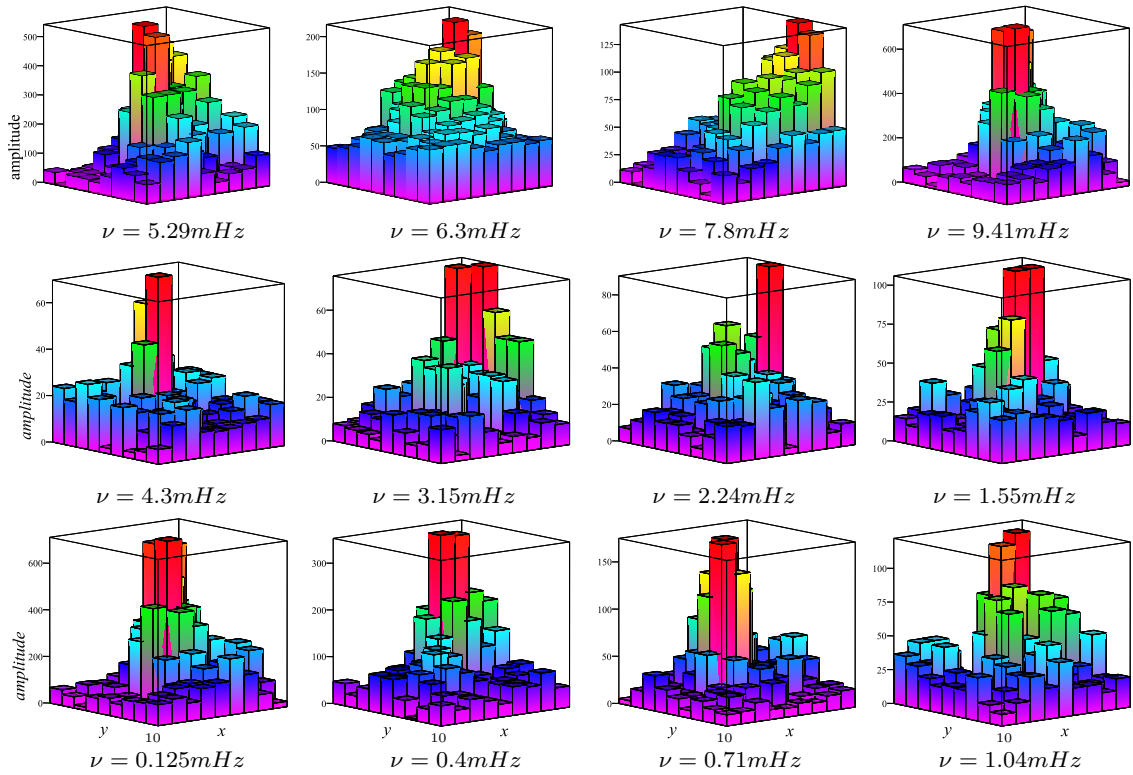}}
\caption{Histograms of  oscillation amplitudes for all pixels of the site under study are shown for a number of frequencies in the entire spectrum range of sunspot oscillations. }\label{fig4}
\end{figure}
Plots of Figure \ref{fig3}a,b plot shows the  oscillation mean amplitude spectrum (blue line) for the entire site (9x9) in the sunspot center. This mean spectrum was obtained by averaging 81 spectra of all the pixels  of the square site shown on Figure \ref{fig1}.  The same plot shows the amplitude spectrum (red line) of the site averaged oscillations that were obtained by averaging oscillations in all the pixels of the investigated site. These two spectra substantially differ from each other both in amplitude and in form. This evidences that the oscillations are not non-local. In case of non-local oscillations, these two spectra should agree with each other. Plots of Figure \ref{fig3}c,d  show (in different colors) the amplitude spectra of nine pixels with coordinates $x=1\ldots 9,y=4$. The averaged oscillation spectrum shown by the red curve in Figure \ref{fig3}a,b is exhibited by the black curve in the Figure \ref{fig3}c,d plot. The number of pixels whose spectra are shown in Figure \ref{fig3}c,d also includes the pixel with coordinates (3,4) where the most-amplitude oscillations are observed. In these two figures, one can see that the oscillation spectra in pixels dramatically differ both from one another, and from the averaged oscillation spectrum of the site. The listed peculiarities of the Figsure \ref{fig3}a,b,c,d spectra are the evidence that the sunspot oscillations are local oscillations, i.e., they represent a set of oscillation cells.
To study the size of the oscillation cells, we built pixel amplitude histograms for fixed frequencies in the entire  spectrum range of sunspot oscillations.  In total, there were built 250 histograms for frequencies that encompass all the sunspot oscillation spectrum, from 0.125 mHz to 10 mHz. We made the histograms for the frequencies that were a certain step from each other in each frequency band. Almost all the histograms indicated the presence of oscillation cells. In total, there were no oscillation cells on several diagrams only. But on these diagrams, there was a very low level of oscillation amplitudes in all the pixels. Figure \ref{fig4} provides the examples of typical oscillation amplitude histograms within the investigated site for the oscillations over the whole oscillation frequency range. In case of three-minute oscillations (top row of histograms in Figure \ref{fig4}), the histograms are built for the frequencies that correspond to the mean amplitude spectrum maxima (Fig. \ref{fig3}a,b). On the histograms in rows 2 and 3 in Figure \ref{fig4}, one can see that the size of the cells appeared about several pixels over the entire investigated frequency range from $\nu=0.125$mHz to $\nu=12$mHz. For the lowest-frequency oscillations (bottom row of histograms in Figure \ref{fig4}), cells become more compact. Also, the ratio of the cell oscillation amplitude to the oscillation amplitudes in pixels surrounding the cell increases notably. The middle row of histograms in Figure \ref{fig4} demonstrates that five-minute oscillations are also concentrated in small cells.

\subsection{Discussion of the sunspot oscillation analysis}
An analysis of observations at high spectral and spatial resolution  reveals the local oscillations in the wide spectrum range ($0.1-10$mHz). These results provide a starting point for constructing a model of local sunspot oscillations. The model is based on two properties of sunspot oscillations. The first property is that fluctuations in the spot consists of individual cells whose size is at the limit of resolution. The second property is that  the spectrum of local oscillations consists of a large number of narrow lines.

\section{Scenarios of local and nonlocal sunspot oscillations}
\citet{Zhugzhda14} analyzed two possible scenarios for the origin of three-minute oscillations in sunspots. Based on this analysis and observation data the model of local three-minute oscillation has been proposed. We generalize the analysis of \citet{Zhugzhda14} to the case of low frequency oscillations.

According to the MAG wave theory (\citealt{Zhugzhda79, Zhugzhda82}) sunspot atmosphere consist of three layers. The weak field approximation is valid in the subphotospheric layers where the magnetic pressure appears much less than the gas one. Slow and fast waves propagate independently from each other and there is no coupling between them in this layer. In this case fast waves are p-modes slightly modified by weak magnetic field. The strong field approximation is valid in the chromosphere and corona where the gas  pressure appears much less than the magnetic pressure one. There is no coupling between slow and fast layers in these layers. Fast and low-frequency slow waves are evanescent ones while high-frequency waves are running ones in under sunspot. The coupling between slow and fast waves is possible in the photospheric layers of sunspot where the magnetic and gas pressure are of the same order.
\begin{figure}%%%%%%%%%%%%%%%%%% FIGURE 5
\centerline{\includegraphics[width=1.0\textwidth,clip=]{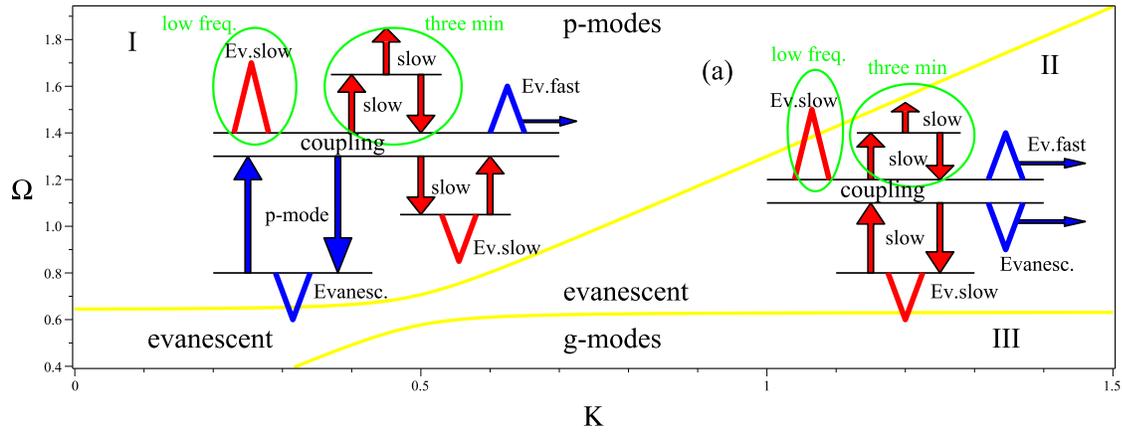}}
\caption{The schemes of two scenarios for MAG wave propagation in a magnetized atmosphere  and diagnostic diagram  for an isothermal atmosphere in terms of dimensionless frequency and horizontal wavenumber $(K=k_\perp H,\Omega=\omega C_S/g)$ are shown. The boundaries between the p-mode  region (I), g-mode region (II), and evanescent waves region are shown by yellow curves.   Red color denotes slow waves, and dark blue signifies fast waves. Arrows mark running waves, and triangles stand for evanescent ones.}\label{fig5}
\end{figure}
It is assumed that the source of sunspot oscillations observed in the photosphere, chromosphere, and corona is located in the subphotospheric layers of the sunspot. Sunspot oscillations can appear due to p-modes or slow waves in subphotospheric layers. These two cases correspond to two scenario of sunspot oscillations. Observations show that there are local sunspot oscillations. \citet{Zhugzhda14} show that local three-minute oscillations appear due to slow waves arriving from subphotosperic layers of sunspot.  The results of data analysis on plots of Figure \ref{fig4} show that local three-minute oscillations are accompanied by low-frequency local oscillations. Besides, there are nonlocal waves, which manifest itself as wave fronts travelling through umbra to penumbra (\citealt{Zirin72, Giovanelli72, Sych14,Su16}).

Figure \ref{fig5} shows two scenarios for both local and nonlocal in sunspots. These scenarios are an extension of scenarios developed by \citet{Zhugzhda14} for three-minute oscillations. It is shown by \citet{Zhugzhda14} that nonlocal three-minute oscillations of sunspot appear due to subphotospheric p-modes while local three-minute oscillations appear due subphotospheric slow waves. Both scenarios include chromospheric resonator and subphotospheric resonators for p-modes and slow modes as it was assumed by \citet{Zhugzhda81}.   A rich spectrum of local oscillations, including a large number of lines,  is impossible to explain by chromospheric resonator since its spectrum consists of just few spectral lines (\citealt{Settele01}). Moreover chromospheric resonator can not explain the occurrence of a sunspot oscillations with periods greater than three minutes. Only subphotospheric resonator for slow waves can be considered as a candidate for explaining of the entire spectrum of local sunspot oscillations.

\section{Local oscillations and Roberts equation}

The further analysis of local slow waves in a sunspot atmosphere is based on the equation for slow waves that was derived by \citet{Roberts06} in limit $k_z/k_\perp\to\infty$  for the case of vertical uniform magnetic field. This equation is true for an arbitrary conductive atmosphere in a homogeneous magnetic field. This is a second-order equation, which means that in this limit there is not a interaction between slow and  fast waves. The Roberts equation for slow waves reads
\begin{eqnarray}
&&C_T^2\frac{{\rm d}^2\xi_z}{{\rm d}z^2}+\gamma g\frac{C_T^4}{C_S^4}\frac{{\rm d}\xi_z}{{\rm d}z}+\left(\omega^2-\frac{C_T^2}{V_A^2}\left(N^2+\frac{g}{H}\frac{C_T^2}{C_S^2}\right)\right)\xi_z=0,
\label{roberts}\\
&&N^2=g\left(\displaystyle\frac{{\rm d}\ln\rho_0(z)}{{\rm d}z}-\frac{g}{C_S^2}\right)=g\left(\frac{(\gamma-1)g}{C_S^2}-\frac{{\rm d}\ln{T}}{{\rm d}z}\right),\label{brunt}
\end{eqnarray}
where $\xi_z$ is the vertical displacement, and $N$ is the Brunt-V\"{a}is\"{a}l\"{a} frequency. The second alternative expression for the Brunt-V\"{a}is\"{a}l\"{a} frequency is written in the form of the Schwarzschield criterion. The square of the Brunt-V\"{a}is\"{a}l\"{a} frequency becomes negative when fulfilling the Schwarzschield criterion for convection occurrence. Thus, the square of the Brunt-V\"{a}is\"{a}l\"{a} frequency is negative in the convection zone. The detailed analysis of Roberts equation can be found in \citet{Zhugzhda14}. Only two formulas we need in the future are listed below.

Robert equation determines the cut-off frequency for the slow waves in a stratified atmosphere with an arbitrary vertical magnetic field. The cutoff frequency is a key parameter which is required in the analysis of wave propagation in a stratified atmosphere. To determine the cut-off frequency, one should transfer from Equation (\ref{roberts}) to an equation without the first derivative. This transition is performed via introducing a new variable
$$\xi_z=\overline{\xi}(z)\exp{\left(-\frac{\gamma g}{2}\int\limits^z\frac{C_T^2}{C_S^4}{\rm dz}\right)}.$$
In terms of this new variable, Equation (\ref{roberts}) comes to the form
\begin{equation}
C_T^2\frac{{\rm d}^2\overline{\xi}(z)}{{\rm d}z^2}+(\omega^2-\Omega_{off}^2)\overline{\xi}(z)=0,\label{nroberts}
\end{equation}
where the so-called cut-off frequency $\Omega_{off}$ is equal to
\begin{equation}
\Omega_{off}^2=\frac{C_T^2}{V_A^2}\left(N^2+\frac{g}{H}\frac{C_T^2}{C_S^2}\right)
+\frac{\gamma^2 g^2}{4}\frac{C_T^6}{C_S^8}+\frac{\gamma gC_T^4}{C_S^4}\left(\displaystyle\frac{{\rm d}\ln{C_T}}{{\rm d}z}-2\frac{{\rm d}\ln{C_S}}{{\rm d}z}\right).\label{cutoff}
\end{equation}
Slow waves with a frequency less than the cut-off frequency are evanescent waves. This formula is the exact one. The cutoff frequency for the isothermal atmosphere permeated by a uniform magnetic field is a very special case of the formula Equation (\ref{cutoff}). The cutoff frequency for the isothermal atmosphere can be applied only to a treatment of the temperature minimum in the sunspot atmosphere.

Equation (\ref{roberts}) is an equation with coefficients depending on the depth $z$, which makes the transition to a dispersion equation impossible. However, for qualitative analysis, one may use approximation of local dispersion approximation when a proper $k_z$ wave number is introduced for each depth in the atmosphere. In this case, Equation (\ref{nroberts}) comes to a local dispersion equation, and it appears possible to determine a local phase and a group velocity for slow waves
\begin{equation}
\omega(k_z)=\sqrt{C_T^2k_z^2+\Omega_{off}^2},\ v_{ph}=\displaystyle{C_T}\left/{\sqrt{1-\frac{\Omega_{off}^2}{\omega^2}}}\right.,\  v_{gr}={C_T}{\sqrt{1-\frac{\Omega_{off}^2}{\omega^2}}}.\label{phvel}
\end{equation}
These formulas for a phase velocity are approximate. In fact, they are more exact, than the oft-used formulas for the phase velocity of slow waves. Slow waves in a strong field are usually considered to propagate with the sound speed, because $C_T\approx C_S$ for $V_A^2\gg C_S^2$. In fact, the phase velocity of slow waves is close to the sound speed only at the  condition $\omega^2\gg\Omega_{off}^2$ , like it follows from Eq.(\ref{phvel}). Just as the phase velocity of slow waves in a weak field is close to the Alfv\`{e}n velocity only at the fulfillment of an additional condition  $\omega^2\gg\Omega_{off}^2\approx N^2$.

Having in hands all the necessary formulas we try to develop a model of local sunspot oscillations.

\section{Occurrence of subphotospric resonator for slow waves}

\begin{figure}%%%%%%%%%%%%%%%%%% FIGURE 6
\centerline{\includegraphics[width=.50\textwidth,clip=]{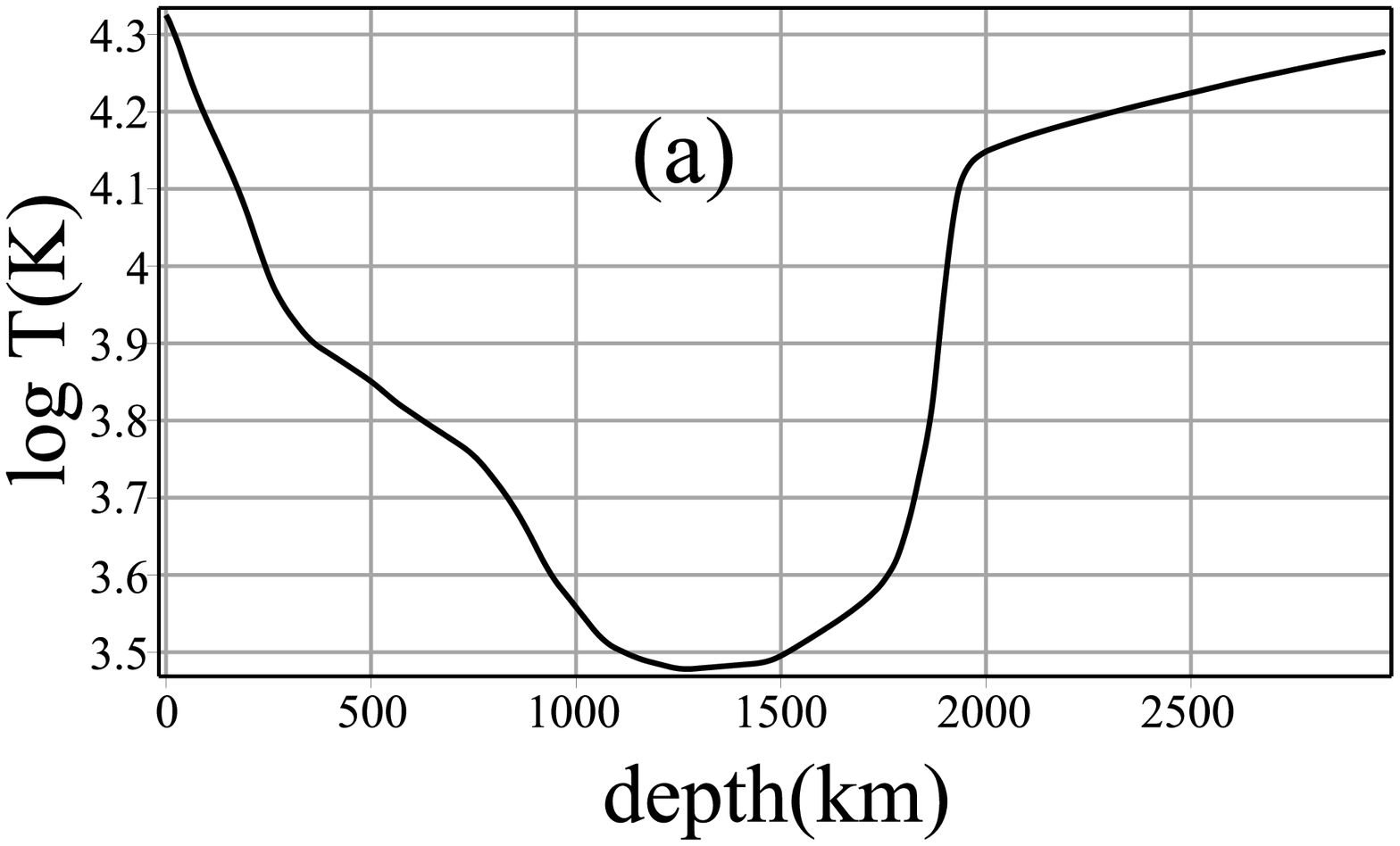}
\includegraphics[width=.50\textwidth,clip=]{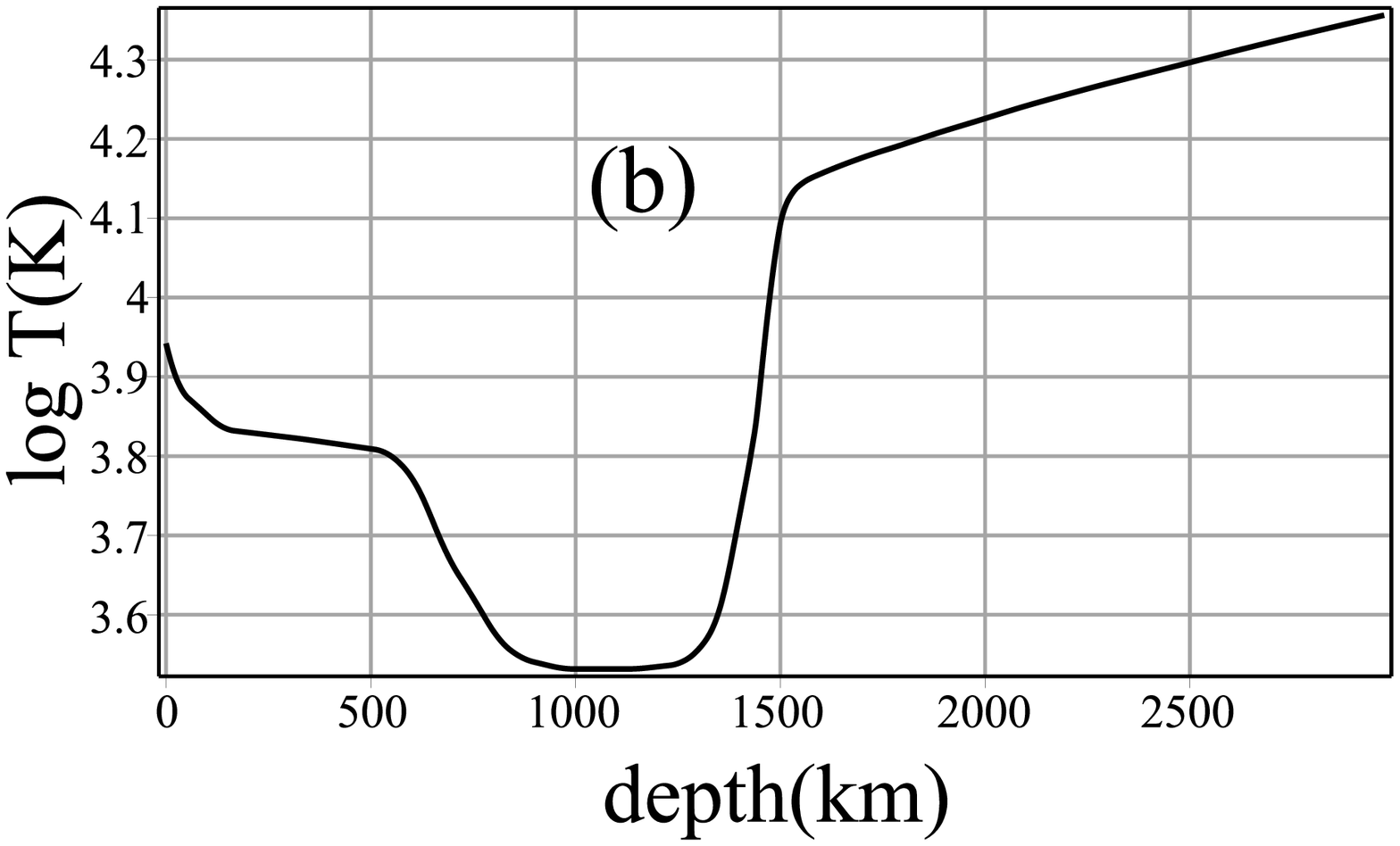}}
\centerline{\hspace{0.5cm}\includegraphics[width=0.50\textwidth,clip=]{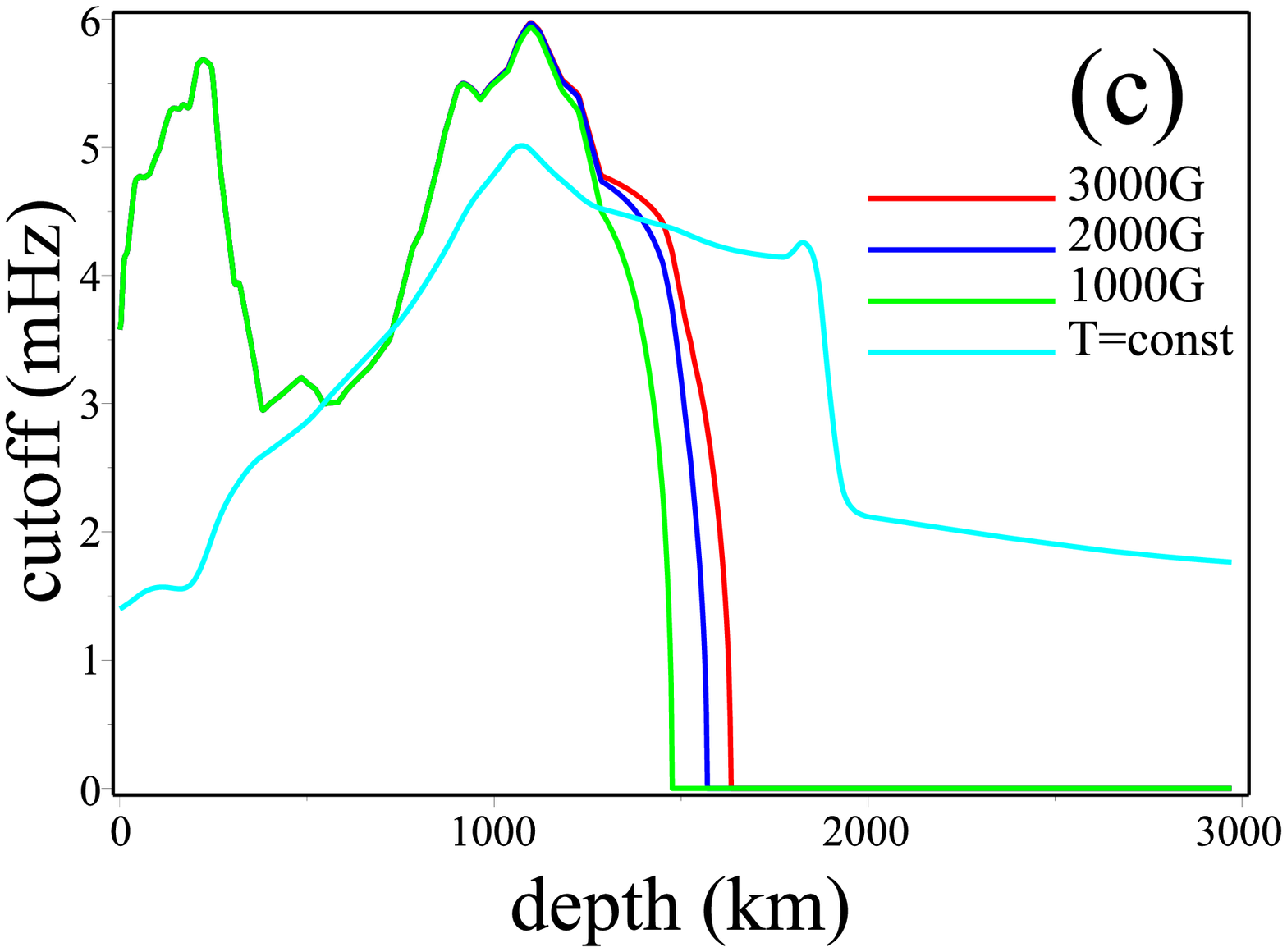}
\includegraphics[width=0.50\textwidth,clip=]{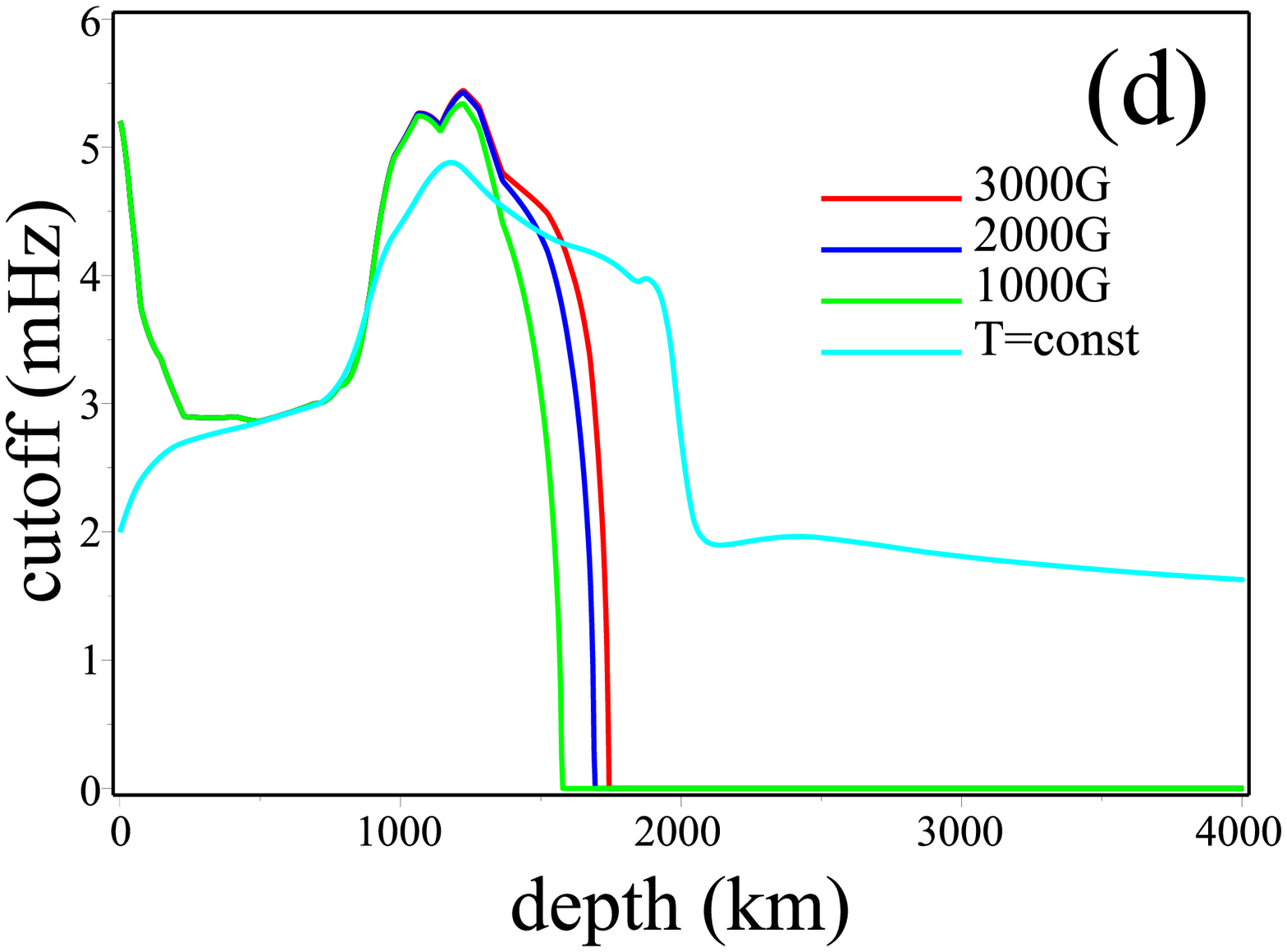}}
\caption{The temperature as function of the depth for the Staude (a) and Maltby (b) sunspot models is shown. The depth is counted from an upper boundary of the chromosphere models. The cut-off frequencies $\Omega_{off}$ as a function of depth for the Staude (c) and Maltby (d) sunspot models are calculated for different values of the sunspot magnetic field $B=3000,2000,1000$Gs. Besides, the slow wave cut-off frequency for the isothermal atmosphere is shown on the plot. }\label{fig6}
\end{figure}

To calculate the model of local sunspot oscillations, we used empirical models of a sunspot atmosphere developed by \citet{Staude81} and \citet{Maltby86}. \citet{Settele01} updated these models and included sunspot subphotospheric layers in them. We also had to update these models a little to enable necessary calculations for the model of local sunspot oscillations. As far as we know \citet{Moradi10}, there are no other models of sunspot that would include the entire sunspot atmosphere, from the chromosphere to the subphotospheric layers, and would contain all the parameters necessary for our calculations. Figsure \ref{fig6}c,d present the results for calculating the cut-off frequency $\Omega_{off}$ (see Eq. (\ref{cutoff})) for the Staude and Maltby models. For comparison, the same plots show the calculations of the cut-off frequency  an isothermal atmosphere in the strong field. Certainly, the approximation of an isothermal atmosphere is not valid for the sunspot photosphere where the strong temperature gradient occurs. But for the temperature minimum, the cut-off frequencies provided by the two formulas are rather close due to a weak temperature gradient. The discussion of reasons of differences between the calculation results for isothermal and non-isothermal atmospheres can be found in \citet{Zhugzhda14}. Plots of cut-off frequency show that the chromospheric resonator occurs. But the chromospheric resonator cannot be responsible for the three-minute oscillations because it cannot provide such a number of spectral lines over the 6-10 MHz frequency range. Calculations by \citet{Settele01} show that the chromospheric resonator leads to the emergence only a few spectral lines (not tens of lines like it occurs in fact) within this frequency range.
Because Figure \ref{fig6}c,d presents real parts of the cut-off frequency, the cut-off frequency under a sunspot  starting with a certain depth turns to zero.  The cut-off frequency in the lower part of the photosphere of the sunspot decreases sharply due to the fall of the Alfven speed. It does mean that slow waves under photospheric layers of sunspot are running waves. This is the confirmation of Zhugzhda's prediction \citet{Zhugzhda84} about the occurrence of the subphotospheric slow-wave resonator.

The occurrence of subphotospheric resonators may lead to the emergence of a spectrum with numerous lines. This is possible only in case of depth-extended resonator with a sufficiently low fundamental resonance frequency. In this case, numerous spectral lines in the subphotospheric resonator are a result of high-harmonic excitation in the subphotospheric resonator.
\begin{figure}%%%%%%%%%%%%%%%%%% FIGURE 7
\centerline{\includegraphics[width=.50\textwidth,clip=]{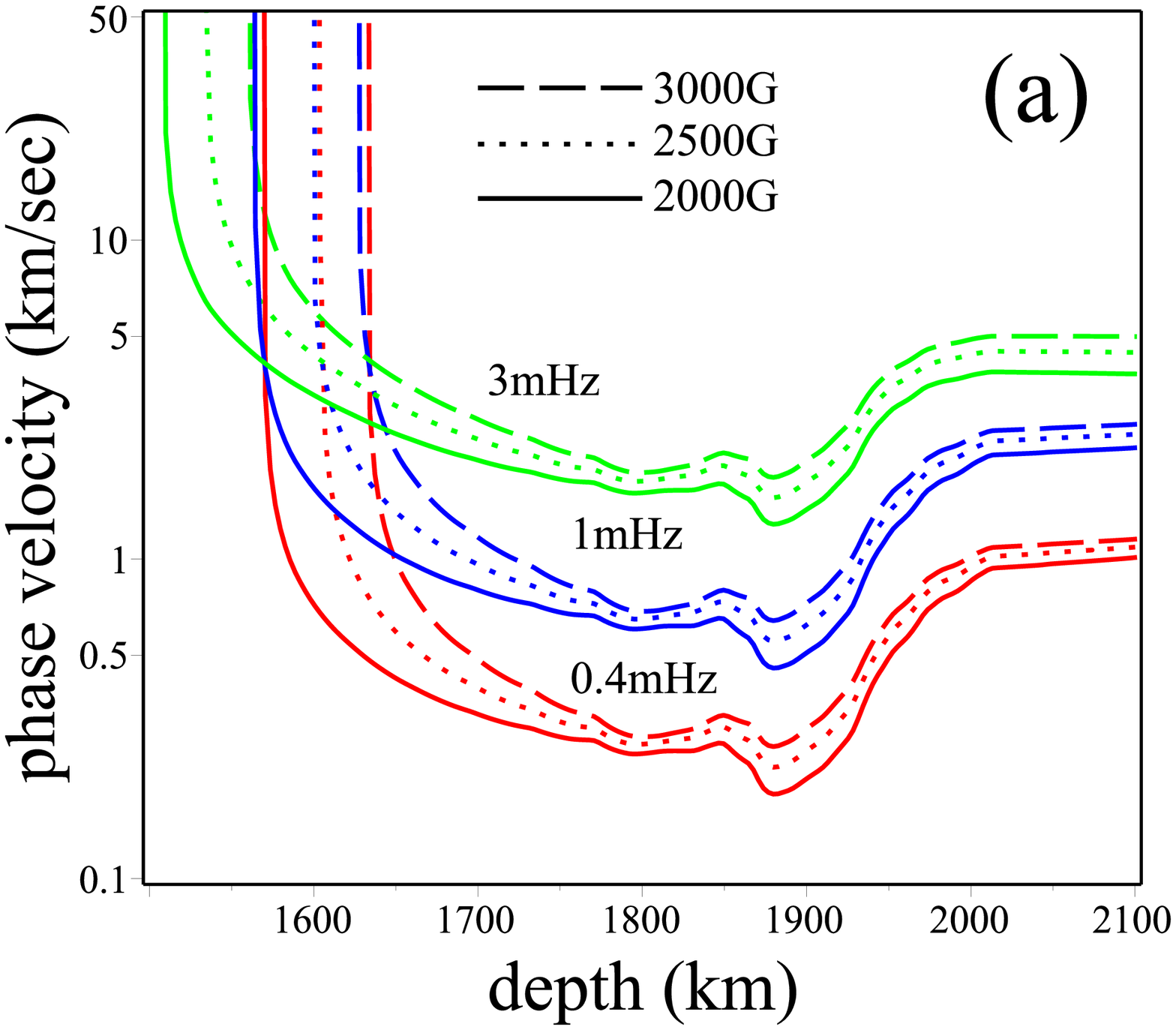}
\includegraphics[width=.50\textwidth,clip=]{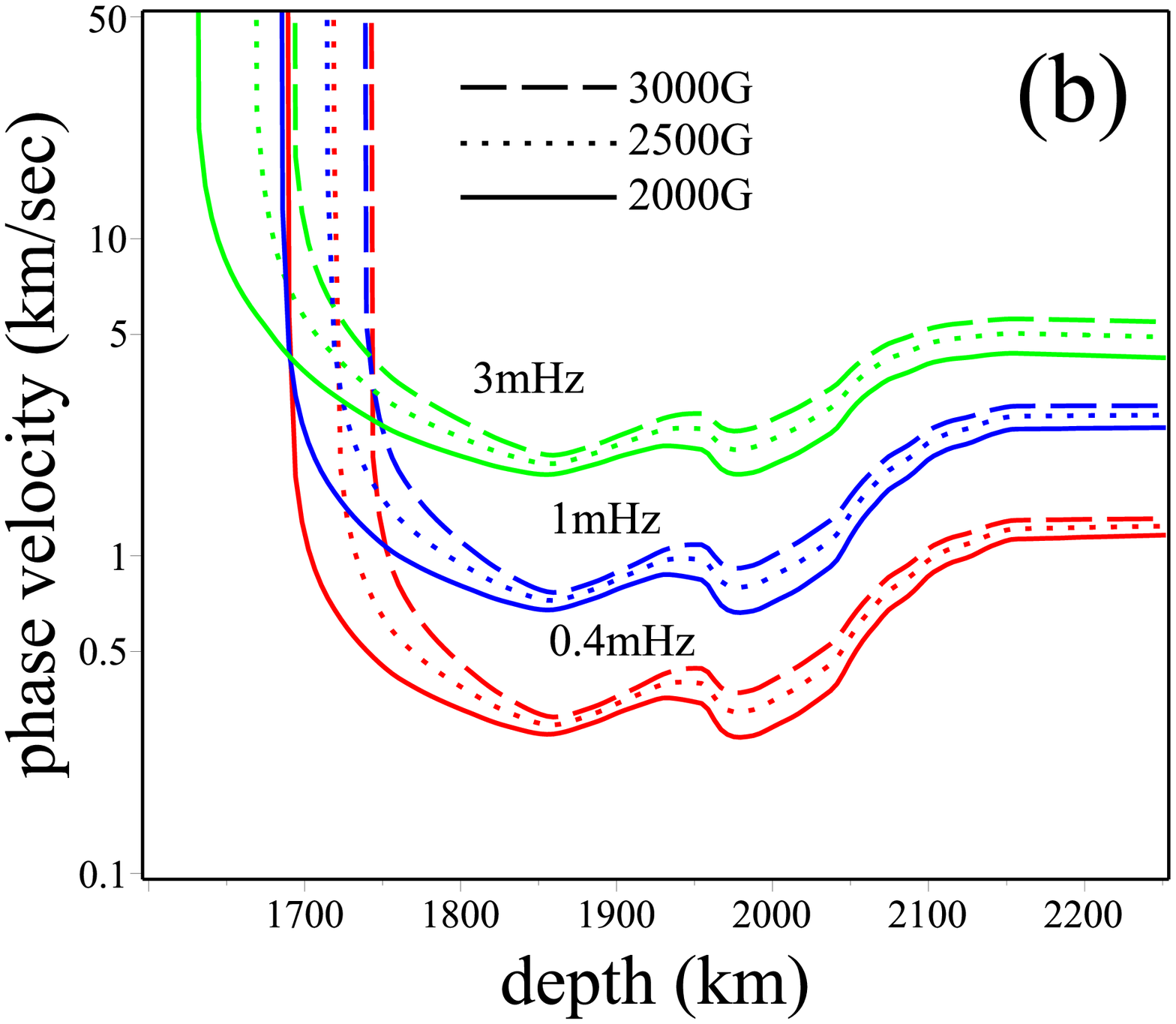}}
\centerline{\includegraphics[width=0.50\textwidth,clip=]{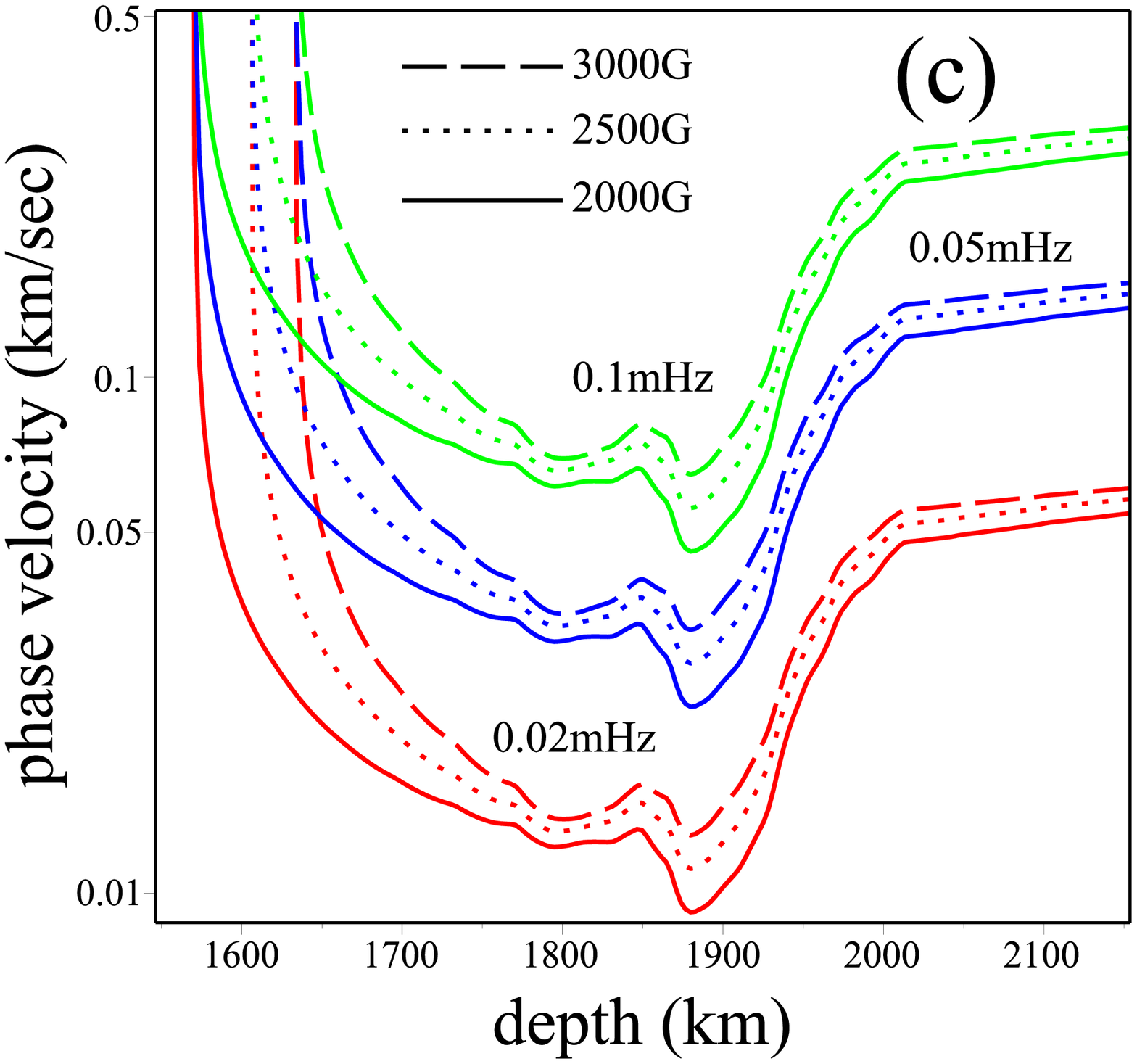}
\includegraphics[width=0.50\textwidth,clip=]{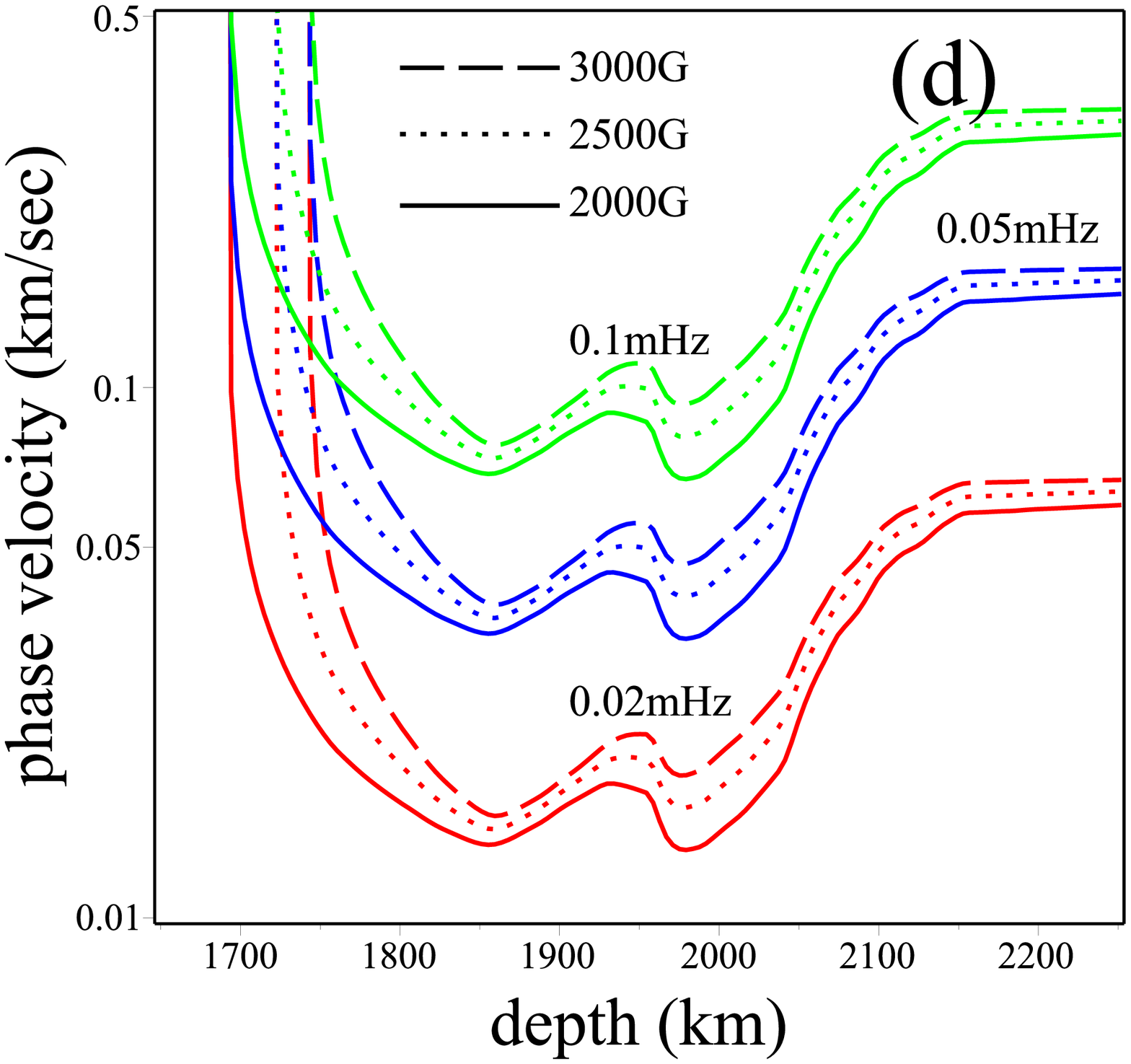}}
\caption{Low-frequency oscillation phase velocity as a function of depth for the Staude (a, c) and Maltby (b, d) models for three values of the magnetic field. }\label{fig7}
\end{figure}
The Figure \ref{fig7} plots present the profiles of the phase velocity for different frequencies and magnetic fields calculated for the Staude and for the Maltby models.
These plots show that local slow waves have strong dispersion. For this reason, the slow-wave resonators significantly differ from those on waves without dispersion. For example, string oscillation eigenfrequencies meet the formula $\nu_n=\nu_0/n$, $n=1,2,3\ldots$, where $\nu_0$ is the fundamental tone, and $\nu_n$ is overtones. In case of the resonator for waves with dispersion, the relation between the fundamental tone and the overtones is non-linear.

There are two main hypotheses for the subsurface structure of sunspots: the monolithic model and the cluster model. Local helioseismology is unable to distinguish between two models (\citealt{Moradi10}). Local oscillations are a good tool to differentiate between models. If the sunspot is a cluster of magnetic flux tubes, some of these tubes may be responsible for local oscillations due to the resonances of the slow waves in them. In the case of monolithic sunspots, convective tongues are candidates for slow mode resonance and local oscillations. The main difference between the magnetic flux tubes of the cluster model and the convective tongues in the monolithic sunspots is the value of the magnetic field. The slow-wave cut-off frequency Equation (\ref{cutoff}) and phase velocity Equation (\ref{phvel}) depend on the magnetic field. Thus, modeling of the sub-photospheric resonator can help to find out the structure of sunspots under the photosphere.

\section{Modeling of subphotospheric resonator}
Modeling of the subphotospheric resonator faces difficulties, since there are no models of magnetic tubes of the cluster sunspot model and convective tongues of the monolithic sunspot model. We were forced to use the Staude and Maltby models for a monolithic sunspot. The description of the subphotospheric layers by these models is quite arbitrary, because of the neglect of convection.

Subphotospheric resonator appears in the lower atmosphere of sunspots, where the slow waves are running ones. Phase velocities (Eq. \ref{phvel}) of traveling slow waves in the lower layers of the sunspots atmosphere are calculated for Staude and Maltby models (\citealt{Settele01}). The results are shown in the plots of Figure \ref{fig7}. At the upper boundary of the subphotospheric resonator cut-off frequency drops to zero (see Fig. \ref{fig6}), and the phase velocity tends to infinity (see Fig. \ref{fig7}). The lower boundary of the resonator is determined by the models of Staude and Maltby. In fact, the slow-wave resonator lower boundary should be located at the depths where the convection starts to work. At the convection, the temperature gradient is close to the adiabatic one, and the square of the Brunt-V\"{a}is\"{a}l\"{a} frequency is close to zero. Thus, the slow wave phase velocity Equation (\ref{phvel}) should increase due to onset of the  convection.  Slow waves should experience a reflection at this level. This is most likely the slow-wave resonator lower boundary. Unfortunately, none of the existing models describes the transition to convection deep under the sunspot. There is an increase in the phase velocity in the middle of the resonator. There is an increase in the phase velocity in the middle of the resonator, which should lead to the splitting of natural frequencies. It makes difficulties to apply standard procedures for eigenproblem. In connection with all the above difficulties, only approximate calculations of the eigenfrequencies of the subphotospheric resonator were carried out to assess the possibility of its occurrence.

The resonance of slow waves in the flux tube in the sunspot atmosphere between depth $z_1$ and $z_2$ is considered. Zero boundary conditions are assumed. To find out the exact  eigenfrequencies Equation (\ref{roberts})  has to be  solved. To find out approximate eigenfrequencies the time $t_n$ as a function of $\omega_n$ was calculated
$$t_n=\int_{z_1}^{z_2}V_{ph}(\omega_n)^{-1}dz,$$
where $V_{ph}$ is defined by Equation (\ref{phvel}). If $n\pi/t_n=\omega_n$, the frequency $\omega_n$ is $n$-th harmonic of resonator. After calculation of the integral eigenfrequencies can be found by intersections on the plot. This method is sufficiently accurate for high harmonics, but rather rough for fundamental one. However, this is not important because of the problems mentioned above, and due to the difficulties of correlation with the observations discussed below.
\begin{figure}%%%%%%%%%%%%%%%%%% FIGURE 8
\centerline{\includegraphics[width=.50\textwidth,clip=]{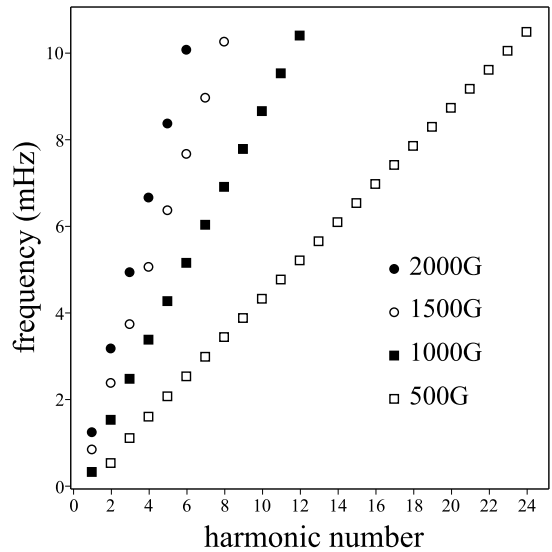}}
\caption{High-frequency resonator eigenfrequencies for different values of the magnetic field calculated for the Staude model.}\label{fig8}
\end{figure}
Figure \ref{fig8} presents the calculations of the subphotospheric resonator eigenfrequencies for the Staude model at different values of the magnetic field. We had to address the magnetic field low values because only in this case the number of harmonics over the three-minute oscillation range appears sufficiently great to correspond to the observational data.

The natural next step after modeling the subphotospheric resonator should be a comparison of the model with the spectrum of oscillations in the sunspot. But, the spectra in Figure \ref{fig2},\ref{fig3} are not at all similar to the spectrum of resonance oscillations consisting of equidistant lines.  The spectrum of sunspot oscillations looks like a noise spectrum without signs of regularity. This is due to the fact that the amplitudes and phases of the resonator harmonics are functions of time because of changes in the size and magnetic field of the convective tongues. We faces the problem to extract from the noisy spectrum of sunspots oscillations the signs of the subphotospheric resonator occurrence. There are the signs of the presence of a resonance in the oscillation spectrum. It is well known that three-minute oscillations consists of a sequence of wave trains. If two lines dominate in the spectrum, then the wave trains will appear because of the beating between the components. If the spectrum consists of equidistant lines due to the subphotospheric resonance, the wave trains have to appear. Despite the fact that the spectrum of the oscillations is similar to the noise spectrum this effect was revealed by \cite{Sych12}. The length of the wave train is determined by the distances between the equidistant lines in the spectrum. In turn, the distances between equidistant lines are equal to the fundamental frequency of the resonance. \citet{Sych12} found out that the characteristic duration of the wave trains of the order of tens of minutes. This is in line with our idea of a low-frequency subphotospheric resonator. To clarify the evidence of the existence of subphotospheric resonance, it is necessary to develop both special methods of spectral analysis and to consider different excitation mechanisms (\citealt{Zhugzhda18}). In addition to excitation of the subphotospheric resonator by turbulent convection, oscillatory convection in a magnetic field can be significant (\citealt{Syrov68}). Oscillatory convection is the instability of the slow wave. However, addressing these issues is beyond the scope of the paper.

\section{Local oscillations and umbral dots}
The local oscillation cell size, according to some data (\citealt{Centeno05}), is estimated as arcsec fractions. At our observation analysis, the local oscillation cell size is at the limit of the instrument spatial resolution. Local sunspot oscillations should obviously be related to a sunspot fine structure. Umbral dots are the only candidate for this role.  The size of umbral dots are of the same order as the size of the local oscillation cells (\citealt{Ebadi17,Serge16}). Besides, the umbral dots oscillate (\citealt{Jess12, Prasad15, Chae17,Ebadi17}). According to the current ideas, umbral dots originate due to convection in the sunspot magnetic field. Most perfect sunspot convection simulations performed by \citet{Schussler06} showed that umbral dots represent  convective tongue that originate as a result of convection in the magnetic field. In our opinion, convective tongues are candidates for the role of a slow-wave resonator.

Despite the obvious shortcomings of our subphotospheric resonator model, we have come to the magnetic field value and resonator size similar to the convective tongue parameters obtained in the simulation   \citet{Schussler06}. No doubt, one needs calculations of eigenfrequencies of convective tongues. Comparative analysis of natural frequencies of convective tongues and spectra of local oscillations in sunspots can be a test of the results of numerical experiments on convection in sunspots. Besides, it is crucial to check, whether the location of local oscillation cells and umbral dots are the same. It should be noted that according to simulations of \citet{Schussler06} the emergence of small-scale upflow tongues start off like oscillatory convection columns below the solar surface but turn into narrow convective tongues driven by the strong radiative cooling around optical depth unity. This is very similar to the oscillatory convection (\citealt{Syrov68}) that occurs due to the instability of slow waves in a stratified atmosphere permeated by a magnetic field.

\section{Local oscillations and wave fronts}
Besides local oscillations, there are also traveling waves that propagate in the form of wavefronts through the umbra into the penumbra  (\citealt{Giovanelli72, Reznikova12, Sych14,Su16}). It is natural to consider these waves as non-local waves excited by a p-modes.  The simulation of this scenario was done by \citet{Felipe17}. They were forced to postulate  a p-mode localized source at some depth under a sunspot.  However, p-modes in the convection zone represent a distributed source.

Thereby, it makes sense to discuss a possibility of exciting the divergent wavefronts by local oscillations. When addressing the two scenarios for the sunspot oscillation excitation, we, indeed, postulated a non-interaction between local and non-local oscillations. In fact, this was based on the assumption that the approximation \citet{Roberts06} is valid in all the sunspot layers.

Slow waves propagate along field lines in the regions of a relatively strong field (the chromosphere and the corona over a sunspot), and in the regions of a relatively weak field (sunspot subphotospheric layers) in accordance with the MAG-wave theory (\citealt{Zhugzhda79, Zhugzhda82}). This means that the propagation of slow waves in adjacent thin flux tubes is independent of each other. Thus, the Roberts approximation is valid in the upper layers (chromosphere and corona), and in the lower layers (subphotosphere of sunspot). However, the validity of the Roberts approximation (\citealt{Roberts06}) in the photosphere region and directly under the latter depends on the magnetic tube thickness.   However, it is unknown how thin must be the magnetic tube for the validity of the approach of Roberts. Nobody has investigated this issue.  The emission of local oscillations from the magnetic flux tube is possible if it is not thin enough at some level on the way through the photosphere. This emission may be responsible for running waves in the sunspots. This problem needs analysis that is beyond the scope of the paper.

\section{Conclution}
Local waves propagate from the sunspot umbra into the corona through the chromosphere (\citealt{Prasad15, Sych12}).  The amplitude of the local oscillation considerably exceeds the amplitude of the nonlocal fluctuations. Moreover the power of local oscillations is underestimated, because the oscillation cell size is, obviously, beyond the present-day instrument spatial resolution (\citealt {Centeno05}). Consequently, local oscillations must be responsible for the dynamics of the chromosphere and the manifestation of chromospheric oscillations in the corona.

It is argued that the local oscillations can be explained in the assumption of the existence of a subphotospheric resonator. So far, no other interpretations have been proposed. Moreover, the proposed model manages to link up the local oscillations with umbral dots and running waves in sunspots. Undoubtedly, the proposed model of local oscillations should be improved and confirmed on the basis of observations.

\begin{acknowledgements}
We are grateful to the referee for helpful and constructive comments and suggestions. The authors are grateful to the SDO/AIA teams for operating the instruments and performing the basic data reduction, and especially, for the open data policy. This work is partially supported by the Ministry of Education and Science of the Russian Federation, the Siberian Branch of the Russian Academy of Sciences (Project II.16.3.2) and by the Program of basic research of the RAS Presidium No.28. The work is carried out as part of Goszadanie 2018, project No. 007-00163-18-00 of 12.01.2018 and supported by the Russian Foundation for Basic Research (RFBR), grant No. 17-52-80064 BRICS-a.
\end{acknowledgements}

\label{lastpage}

\end{document}